\documentclass[]{pasj01}
\Received{$\langle$09-Apr-2015$\rangle$}
\Accepted{$\langle$12-May-2015$\rangle$}
\Published{$\langle$publication date$\rangle$}

\begin{document}
\SetRunningHead{Baba et al.}{Radial distributions of arm-gas offsets}

\title{Radial distributions of arm-gas offsets as an observational test of spiral theories}

%%% begin:list of authors
% Do NOT capitalize all letters in "textsc".
\author{Junichi \textsc{Baba}$^1$,
Kana \textsc{Morokuma-Matsui}$^{2}$, and Fumi \textsc{Egusa}$^{2,3}$}
\affil{$^1$ Earth-Life Science Institute, Tokyo Institute of Technology, Ookayama, Meguro, Tokyo 152--8551, Japan.}
%\affil{Nobeyama Radio Observatory, 462-2 Nobeyama, Minamimaki-mura, Minamisaku-gun, Nagano 384--1305, Japan} 
\affil{$^2$ Chile Observatory, National Astronomical Observatory of Japan, 2-21-1 Osawa, Mitaka, Tokyo 181--8588, Japan.} 
\affil{$^3$ Institute of Space and Astronautical Science, Japan Aerospace Exploration Agency, Sagamihara, Kanagawa 252--5210, Japan}
\email{babajn@elsi.jp; babajn2000@gmail.com}

\KeyWords{
%xxxx:xxxx ......
    Galaxies: spiral ---
    Galaxies: kinematics and dynamics ---
    ISM: kinematics and dynamics  ---
    method: numerical
} %Do NOT move this preamble from here!

\maketitle

\begin{abstract}
Theories of stellar spiral arms in disk galaxies can be grouped into two classes based on the longevity of a spiral arm.
Although the quasi-stationary density wave theory supposes that spirals are rigidly-rotating, long-lived patterns, 
the dynamic spiral theory predicts that spirals are differentially-rotating, transient, recurrent patterns. 
In order to distinguish between the two spiral models from observations, we performed hydrodynamic simulations 
with steady and dynamic spiral models. 
Hydrodynamics simulations in steady spiral models demonstrated that 
the dust lane locations relative to the stellar spiral arms (hereafter, arm-gas offsets) depend on radius, 
regardless of the strength and pitch angle of the spiral and the model of the inter-stellar medium (ISM). 
In contrast, we found that the dynamic spiral models show no systematic radial dependence of the arm-gas offsets. 
The arm-gas offset radial profile method, together with the other test methods,
will help us to distinguish between the two spiral models in observed spiral galaxies. 
\end{abstract}

\section{Introduction}

Spiral structures in galaxies have been interpreted as ``standing waves'' in a stellar disk 
(the so-called quasi-stationary density wave hypothesis; \cite{LinShu1964,BertinLin1996}).  
The density waves generate spiral perturbations and affect on the gas flows.
When the gas inside a co-rotation radius overtakes a spiral arm as it moves around a galactic disk, 
the gas is expected to form a standing shock, called a ``galactic shock'', at the upstream side of the spiral arm
(e.g., \cite{Fujimoto1968,Roberts1969,Shu+1973}).
Since such a galactic shock strongly compresses the gas, the galactic spiral dust-lanes and 
the associated star formation around spiral arms have since been regarded as consequences of the galactic shocks.
The galactic shock hypothesis qualitatively explains the CO-H$\alpha$ offset distributions 
observed in some spiral galaxies such as M51 (e.g., \cite{Egusa+2009,Louie+2013}). 
However, it is still unclear whether a spiral arm rotates with a single pattern speed 
for several rotational periods (i.e., $\sim$ 1 Gyr), as predicted by the quasi-stationary density wave hypothesis.

On the other hand, a transient spiral hypothesis was proposed in the 1960s \citep{GoldreichLynden-Bell1965,JulianToomre1966}.
This hypothesis has been extended to a transient recurrent spiral  (hereafter, ``dynamic spiral'') hypothesis by 
$N$-body simulations of stellar disks (see a review by \cite{DobbsBaba2014}); 
the amplitudes of stellar spiral arms change on the time scale of galactic rotation or even less
(i.e, a few hundreds of Myrs; \cite{SellwoodCarlberg1984,Baba+2009,Fujii+2011,Sellwood2011,Grand+2012a,Baba+2013,D'Onghia+2013,Pettitt+2015}),
and the spiral at any given radius co-rotates with materials \citep{Wada+2011,Grand+2012a,Baba+2013,Kawata+2014}. 
\citet{Fujii+2011} revealed a self-regulating mechanism that maintains multi-arm spiral features for at least 10 Gyrs.
The ``dynamic equilibrium'' nature of stellar spiral arms can be attributed to their co-rotational nature \citep{Baba+2013}.
Furthermore, \citet{DobbsBonnell2008} and Wada, Baba \& Saitoh (\yearcite{Wada+2011}) 
performed hydrodynamic simulations of the gas flows in dynamic spirals, and showed that gas does not flow through a spiral arm, 
but rather effectively falls into the spiral potential minimum from both sides of the arm (called ``large-scale colliding flows'').
The dynamic spiral hypothesis is also supported by observations (e.g., \cite{Foyle+2011,Ferreras+2012}).

Some observational tests for discriminating between quasi-stationary density waves (i.e., steady spirals) 
and dynamic spirals have been proposed (see a review by \cite{DobbsBaba2014}).
In this paper, we propose a new indicator to discriminate between the two spiral models 
based on the radial distributions of arm-gas offset angles.
The dust lanes locations relative to a stellar spiral arm depend on the gas flow in the spiral galaxy.
In the following sections, 
we demonstrate that the arm-gas offset distributions show different radial dependences between 
hydrodynamic simulations of steady spiral and dynamic spiral models.
This difference suggests that measuring the radial distribution of the arm-gas offset angles 
can be a new observational means of distinguishing between the two spiral models.

\section{Models and methods}

We performed hydrodynamic simulations in rigidly rotating spiral potentials and $N$-body/hydrodynamic simulations 
of stellar and gas disks with the $N$-body/smoothed particle hydrodynamics (SPH) simulation code {\tt ASURA-2}
(\cite{SaitohMakino2009},\yearcite{SaitohMakino2010}).

\subsection{Steady spiral models}

To simulate the gas flows in spiral galaxies with stationary density waves, 
we impose the rigidly rotating spiral potentials into the axisymmetric potential.
The gravitational potential of the model galaxy consists of an axisymmetric and a non-axisymmetric (i.e., spiral) parts. 
The static axisymmetric part of the gravitational potential comprises of a stellar disk, a spherical bulge and halo.
The initial radial distribution of the gas follows an exponential profile with a scale length of $\sim 11$ kpc,
which is motivated by the observations of the Milky Way galaxy (e.g., \cite{BigielBlitz2012}).
Figure \ref{fig:RotationCurve} shows the circular velocity curves of each component.

The gravitational potential of the spiral arm is given by 
\begin{eqnarray}
\Phi_{\rm sp}(R,\phi,z;t) 
&=& A(R,z) \frac{z_{\rm 0}}{\sqrt{z^2+z_{\rm 0}^2}} \nonumber \\
&\times& 
\cos \left[m \left(\phi - \Omega_{\rm p}t + \cot i_{\rm sp}\ln\frac{R}{R_{\rm 0}}\right) \right],
\end{eqnarray}
where $A$, $m$, $i_{\rm sp}$, $\Omega_{\rm p}$, and $z_{\rm 0}$ 
are the amplitude of the spiral potential, the number of stellar spiral arms, 
the pitch angle, the pattern speed, and the scale-height, respectively. 
Note that $\Phi_{\rm sp}$ decreases from $R=1$ kpc exponentially to be $\Phi_{\rm sp} = 0$ at the center.
In this study, we set $m =2$, $z_0 = 100$ pc, and $R_0 = 1$ kpc. 
We adopt a typical value of the pattern speeds of spiral arms in Milky Way galaxy, 
$\Omega_{\rm p} \simeq 23~\rm km~s^{-1}~kpc^{-1}$ (i.e., $R_{\rm CR} = 10$ kpc), 
although they have a wide range of $18$ -- $30~\rm km~s^{-1}~kpc^{-1}$ \citep{Gerhard2011}.
The amplitude of the spiral potential is controlled by the dimensionless parameter 
\begin{equation}
\mathcal{F} \equiv \frac{m|A|}{|\Phi_{\rm 0}| \sin i_{\rm sp}},
\end{equation}
where $\Phi_{\rm 0}(R,z)$ is the axisymmetric potential.
This dimensionless parameter measures the gravitational force due to the spiral arms
in the direction perpendicular to the arms relative to the radial force from 
the background axisymmetric potential \citep{Shu+1973}.

In the steady spiral runs, we perform two series of runs with the different ISM models: 
the first series is hydrodynamic simulations of an isothermal, non-self-gravitating gas.
No star formation and stellar feedback are considered. 
The temperature of the gas is assumed to $10^4$ K to effectively include a contribution of turbulent motions.
This ISM model corresponds to that assumed in the classical galactic shock theory (e.g., \cite{Roberts1969}).
The second series is hydrodynamic simulations with a more realistic ISM model,
which includes the self-gravity, radiative cooling for a wide temperature range of $20~{\rm K} < T < 10^8~{\rm K}$ \citep{Wada+2009}, 
heating due to far-ultraviolet radiation \citep{GerritsenIcke1997}, star formation and supernova feedback \citep{Saitoh+2008}.
The self-gravity is calculated by the Tree with GRAPE method with a software emulator of GRAPE, Phantom-GRAPE \citep{Tanikawa+2013}. 
Model parameters are summarized in Table \ref{tbl:ModelParameters}.
The simulations are performed in a frame co-rotating with the spiral arms until $t=350$ Myr.
To avoid strong transients in the gas flows caused by a sudden introduction of the spiral potential,
we increase its amplitude linearly from $t=100$ Myr to $200$ Myr.

\subsection{Dynamic spiral models}

\begin{figure}
\begin{center}
\includegraphics[width=0.36\textwidth]{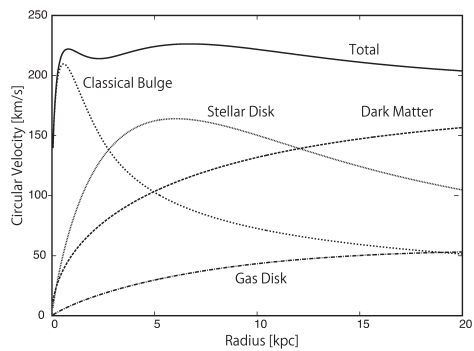}
\caption{Circular velocity curves of each component.
In the cases of the rigidly rotating spiral arms, all components are assumed to be external potentials (section 3.1).
For the case of the dynamically evolving barred-spiral, only the dark matter halo is treated as an external potential,
but other components are solved by the $N$-body/SPH methods (section 3.2)}	
\label{fig:RotationCurve}
\end{center}
\end{figure}

We performed three-dimensional, high-resolution $N$-body/SPH simulations of a Milky Way-like galaxy model.
An initial axisymmetric model of stellar and gaseous disks, a classical bulge embedded in a dark matter halo 
has the circular velocity curves of each component shown in Figure \ref{fig:RotationCurve}.
We take into account the self-gravity of the gas, cooling, heating, star formation and supernova feedback (section 2.1).
In this model, the stellar bar is developed by a bar instability \citep{OstrikerPeebles1973,Efstathiou+1982}
at $t \simeq 1.5$ Gyr.
In this paper, we focus on the gas flows and spatial distributions in simulated 
unbarred ($t \simeq 1$ Gyr) and barred ($t \simeq 2.5$ Gyr) spiral galaxeis.  
Detailed studies on dynamics of stars, as well as structures of the ISM/clouds, will be 
separately discussed in forthcoming papers (J. Baba et al., in preparation). 

\begin{table}[htdp]
\caption{Model parameters for the steady spiral models and properties of spiral arms in dynamic spiral models.
Note that the `multiphase' ISM model includes cooling, heating, star formation and supernova feedback
(See section 2 for details).
For comparison, typical values of $i_{\rm sp}$ and $\mathcal{F}$ for dynamic spiral models are presented. 
}
\begin{center}
\begin{tabular}{lcccc}
Models		& $i_{\rm sp}$ 	& $\mathcal{F}$	& gas self 	& ISM	\\
			& [deg.]		& [\%]			& -gravity	& model	\\
\hline
{\it steady spirals}\\
~Si10F02iso	& 10			& 2				& no		& isothermal \\
~Si10F05iso	& 10			& 5				& no		& isothermal \\
~Si20F05iso	& 20			& 5				& no		& isothermal \\
~Si20F05msg	& 20			& 5				& yes	& multiphase \\ % cooling, heating, SF, FB \\
\hline
{\it dynamic spirals}\\
~DynUnBar	& $\sim 25$	& $\lesssim$ 5		& yes	& multiphase \\ % cooling, heating, SF, FB \\
~DynBar		& $\sim 30$	& $\lesssim$ 8		& yes	& multiphase \\ % cooling, heating, SF, FB \\
\end{tabular}
\end{center}
\label{tbl:ModelParameters}
\end{table}%

\section{Results}
\subsection{Arm-gas offset angles in steady spirals}

\begin{figure*}
\begin{center}
\includegraphics[width=0.95\textwidth]{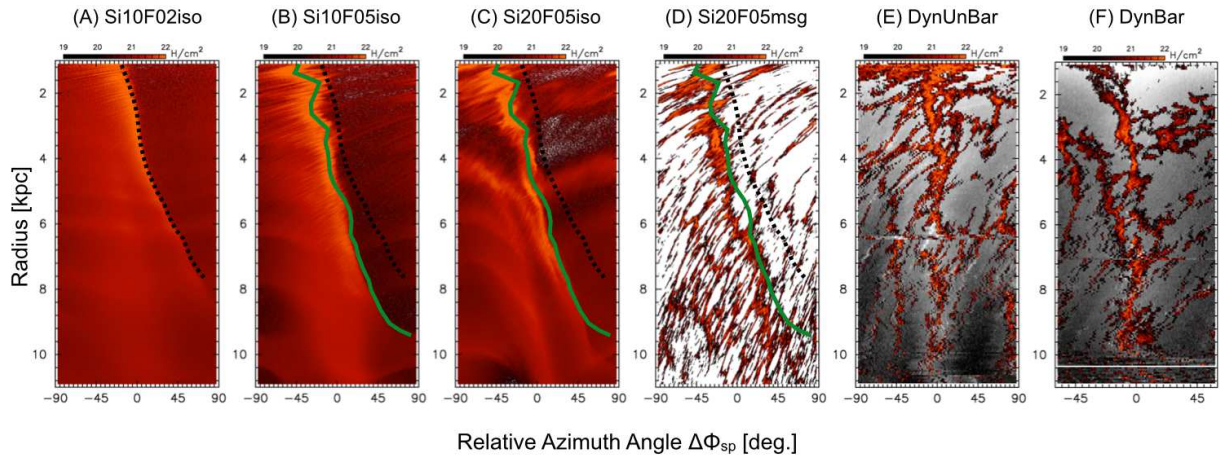}
\caption{
Panels (A)--(D): 
The azimuth angle of the gas relative to the potential minimum ($\Delta \phi_{\rm sp}$)
as functions of radius for models with {\it steady} spirals at $t = 350$ Myr.
The potential minimum of each radius is located at $\Delta \phi_{\rm sp} = 0$. 
The negative value of an offset angle ($\Delta \phi_{\rm sp}<0$) 
corresponds to a leading side (i.e., downstream inside the co-rotating radius).
The dotted (black) lines indicate the locations of shocks in the Si10F02iso model (panel A). 
The solid (green) lines indicate the locations of shocks in the Si10F05iso model (panel B). 
Panels (E) and (F): Same as the panels (A)--(D), but for {\it dynamic} spiral models at $t = 1.12$ Gyr (panel E) and $2.55$ Gyr (panel F). 
The model galaxy is an `unbarred' grand-design spiral at $t < 1.5$ Gyr, 
but is a `barred' grand-design spiral after then (J. Baba et al. in preparation).
The locations of  the grand-design spirals are defined by the phases of the dominant Fourier modes (i.e., $m=2$ or 3).
The background gray-scale map presents a stellar density. 
The grand-design spiral is located at $\Delta \phi_{\rm sp} = 0$.
%At $t=2.55$ Gyr (panel F), a stellar bar is seen at $R < 3$ kpc.
}	
\label{fig:Offset}
\end{center}
\end{figure*}

Our results show that the locations of galactic shocks strongly depend on radius.  
Figure \ref{fig:Offset} shows the radial distributions of the offset angle of the dense gas relative to the spiral potential 
(hereafter, the arm-gas offsets). In the tightly winding, weak spiral case (Si10F02iso model; panel A of Figure \ref{fig:Offset}), 
the galactic shock is located on the upstream side except for $R<2$ kpc. 
This is consistent with the predictions by the classical galactic shock theory \citep{Roberts1969}.
However, the galactic shock is away from the potential minimum and moves upstream 
if it is near the co-rotation radius ($R_{\rm cr} = 10$ kpc). 
This is because the Mach number is a function of radius (e.g., \cite{Shu+1973}).
Such radial dependence is consistent with \citet{GittinsClarke2004}, who found that the pitch angle of gaseous arms 
is smaller than that of stellar arms (see also \cite{KimKim2014}).

%Radially dependent 
The radial trend of arm-gas offset angles in steady spirals depends qualitatively
on neither the strength nor the pitch angle of the spiral.
If we compare results with the different $\mathcal{F}$ models (panels A and B in Figure \ref{fig:Offset}), 
the galactic shock locations in Si10F05iso model are shifted farther downstream (see also \cite{Woodward1975}).
Solid green lines in panels B and C in Figure \ref{fig:Offset} indicate the locations of galactic shocks in the Si10F05iso model.
The locations of galactic shocks in both models are almost the same. 
This is a natural consequence of the definition of $\mathcal{F}$.
Note that the shocked layers in spiral potentials are not always dynamically stable (so-called wiggle instability).
This is consistent with the previous time-dependent, multi-dimensional hydrodynamic simulations 
(e.g., \cite{WadaKoda2004,ShettyOstriker2006,DobbsBonnell2006,Kim+2014}).
The physical origin of wiggle instabilities is a standing problem\footnote{
\citet{HanawaKikuchi2012} pointed out that the wiggle instability (WI) may be due entirely to 
numerical artifacts arising from the inability to resolve a shock inclined to numerical grids.
However, \citet{Kim+2014} found that the WI is a physical phenomenon and originating from 
the generation of potential vorticity at a deformed shock front, 
rather than Kelvin--Helmholtz instabilities as proposed by \citet{WadaKoda2004}. 
We do not aim to investigate the physical origins here.
See these papers for more details.
}, 
although the wiggle instabilities result in forming the substructures such as spurs in the inter-arm regions.

The steady spiral models with more realistic ISM models also show the similar radially dependent arm-gas offset angles.
In panel (D) of Figure \ref{fig:Offset}, we show the radial distribution of the arm-gas offsets for Si20F05msg model,
which include self-gravity of the gas, radiative cooling, heating and stellar feedback.
As suggested by \citet{Lee2014}, the self-gravity of the gas may shift 
the shock location downstream (\cite{Sawa1977,Wada2008}), 
but a clear radial dependence of the arm-gas offset is still seen in Si20F05msg model.
This suggests that an arm-gas offset distribution does not depend on ISM modeling, 
but is essentially determined by dynamical properties of a spiral arm.

\subsection{Arm-gas offset angles in dynamic spirals}

\begin{figure}
\begin{center}
\includegraphics[width=0.48\textwidth]{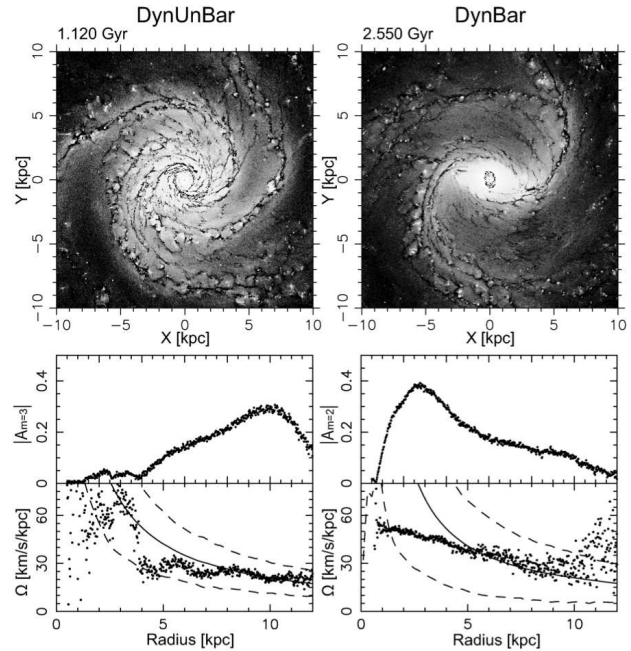}
\caption{
Upper: $B$-band brightness distributions of stars (gray) in {\it dynamic} spiral models 
at $t = 1.12$ Gyr (left) and $2.55$ Gyr (right). 
The dust extinction for the $B$-band map is estimated by multiplying a factor of $e^{-\tau_B}$, 
where the optical depth $\tau_B$ is calculated from the absorption cross section 
$\sigma_B = 6 \times 10^{-22}~\rm cm^2$ and the total hydrogen column density $\Sigma_{\rm H}$. 
Lower: Radial profiles of the Fourier amplitudes and angular phase speeds of the dominant modes ($m=3$ and 2)
at $t = 1.12$ Gyr (left) and $2.55$ Gyr (right), respectively.
Solid line indicates a  galactic rotation frequency $\Omega$, and dashed lines indicate 
$\Omega \pm \kappa/m$, where $\kappa$ is an epicyclic frequency.
The angular speeds are not well determined around radii with weak amplitudes.
}	
\label{fig:DynamicSpiralPatternSpeed}
\end{center}
\end{figure}

We investigate the dust-lane locations in dynamically evolving spiral arms.
Figure \ref{fig:DynamicSpiralPatternSpeed} shows the $B$-band brightness distributions of stars with dust extinction 
at $t=1.12$ Gyr and $t=2.55$ Gyr. The model galaxy is an `unbarred' grand-design spiral at $t < 1.5$ Gyr, 
and then becomes a `barred' grand-design spiral. At $t=2.55$ Gyr, a stellar bar is seen at $R < 3$ kpc. 
Properties of spirals are analyzed by using the 1D Fourier decomposition method as follows:
\begin{eqnarray}
A_m(R) = \int_{-\pi}^{\pi} \frac{\mu(R,\phi)}{\bar{\mu}(R)} e^{-im\phi} d\phi,
\label{eq:FFT}
\end{eqnarray}
where $\mu(R,\phi)$ is the stellar mass distribution and 
$\bar{\mu}(R)$ is the azimuthally averaged surface stellar density at radius $R$, 
and $A_m(R)$ denotes a complex amplitude of the $m$-th mode, respectively.
We analyze angular positions and angular phase speeds of spiral arms 
with the Fourier density peak method \citep{Wada+2011,Roca-Fabrega+2013}.

The angular phase speeds of spiral arms in dynamic spiral models follow the galactic rotation at almost all radii
or radially decrease (bottom panels of Figure \ref{fig:DynamicSpiralPatternSpeed}). 
This is similar to the simulated multi-armed spirals (e.g., \cite{Baba+2013}), 
as well as the simulated barred spirals (e.g., \cite{Grand+2012a}),
which have the radially dependent rotation of spiral arms, but slightly faster than the galactic rotation.
This suggests that spiral arms even in barred galaxies could not be rigid-body rotating patterns predicted 
by the quasi-stationary density wave theory.

In contrast to the steady spiral models, we can clearly see that the dust lanes are 
along with stellar spiral arms in the dynamic spiral model (top panels of Figure \ref{fig:DynamicSpiralPatternSpeed}).
As shown in the panels (E) and (F) of Figure \ref{fig:Offset}, 
the radial distributions of the arm-gas offset angle show no clear radial dependence in dynamic spiral models.
This is because the gas does not flow through the spiral arm, and
the gaseous arm remains until the stellar arm disperses (see also  \cite{DobbsBonnell2008,Wada+2011}).

\section{Discussion and Summary}

In this study, we performed numerical simulations with steady and dynamic spiral models and 
compared the locations of dense gas around spiral arms.
The steady spiral models predict that the dust lanes are located downstream of the stellar spiral in the inner regions 
and shift upstream of the spiral in the outer regions. 
In contrast, the dynamic spiral models show no systematic radial dependence of the arm-gas offsets,
since the gas motions follow large-scale colliding flows to form spiral arms.
Thus, our results suggest that a radial profile of arm-gas offset angle can be used to distinguish between the two spiral models. 

To measure the radial distributions of arm-gas offsets, spatially-resolved wide-field maps of old stars and cold gas are required. 
\citet{Kendall+2011} used stellar mass density maps and gas shocks traced by the {\it Spitzer} Infrared Array Camera (IRAC) 8 $\mu$m data,
and investigated arm-gas offset profiles of nearby spiral galaxies.
Forthcoming integral field spectroscopic surveys such as the MaNGA Survey \citep{Bundy+2015} and wide-field mapping by ALMA will 
provide us accurate spatially resolved stellar and gas maps of nearby galaxies, 
and will facilitate discrimination between the two spiral theories.

However we should note that the steady and dynamic models are not the only options. 
For example, \citet{Dobbs+2010} presented a tidally-induced spiral model 
in which the pattern speed is slower than the galactic rotation everywhere but decreases as a function of radius. 
In order to determine the nature of spiral structures in a spiral galaxy, a single test is not enough. 
Other methods, such as CO-H$\alpha$ offset measurement \citep{Egusa+2009} 
and velocity-field modeling \citep{KunoNakai1997,Miyamoto+2014},
are complementary to the test proposed in this paper.

\bigskip

The authors thank the anonymous referee for careful reading.
%and constructive criticism which has significantly improved the paper. 
JB thank also Takayuki R. Saitoh, Jin Koda, and Keiichi Wada for their constructive comments.
Calculations, numerical analyses and visualization were carried out on Cray XT4, XC30,
and computers at Center for Computational Astrophysics, National Astronomical Observatory of Japan.  
JB was supported by HPCI Strategic Program Field 5 ``The origin of matter and the universe''
and JSPS Grant-in-Aid for Young Scientists (B) Grand Number 26800099.

%\bibliographystyle{apj}
%\bibliography{ms}

\end{document}